\documentstyle[12pt,aaspp4,epsf]{article}
\begin{document}
\title{Neutrino Induced Electrostatic Waves in Degenerate Electron Plasmas}
\author{J. Martin Laming\\
SFA, Inc., 1401 McCormick Drive, Largo, MD 20774\\
(also Naval Research Laboratory, Code 7674L, Washington DC 20375)\\
e-mail jlaming@ssd5.nrl.navy.mil}
\begin{abstract}
The plasma theory of electrostatic waves generated collisionlessly when
neutrinos pass through degenerate electrons is developed. It is consistent with
quantum statistical field theory in the degenerate limit. The reactive case can also be treated. For neutrino distributions considered herein, neither case
develops significantly during e.g. the neutrino burst in a supernova explosion.
\end{abstract}

\section{Introduction}
A number of recent papers have speculated on the collisionless heating of
thermal plasma behind a Type II supernova shock by the intense flux of
neutrinos emitted from the supernova core. The idea is attractive since
the total energy radiated by a supernova in neutrinos (and antineutrinos,
$\sim 10^{53} - 10^{54}$ erg)
far exceeds that in the thermal plasma ($\sim 10^{51}$ ergs), and so a small
transfer of energy from neutrinos to electrons and nuclei, in excess of that
already accounted for by collisional processes alone, could go a long way
to solving what has become known as the Type II supernova problem, i.e., that
in models, the supernova shock following the core collapse and bounce 
loses energy to nuclear dissociation, stalls and  eventually collapses back.
Such an event would produce neutrinos, but not the optical
emission for which supernovae are famed.

Bingham and coworkers [\cite{bingham94}]
derived a growth rate for the neutrino analog of the
reactive two stream instability
in hot ($\sim 3\times 10^9$ K) plasma, treating the
coupling between the neutrinos and the plasma by an effective potential
$\sqrt{2}G_Fn_e$, following [\cite{bethe86}],
where $G_F$ is the Fermi weak interaction constant and
$n_e$ is the electron density. A number of the assumptions in this paper were
criticized by Hardy \& Melrose, who used an approach based on
quantum statistical field theory to derive a growth rate for the kinetic
instability. Aside from rather elementary assumptions like the use by 
[\cite{bingham94}] of a
monoenergetic neutrino beam rather than a Fermi-Dirac distribution, and the
treatment of the thermal plasma with a cold electron dielectric tensor
[\cite{hardy97}]
show that their result for the growth rate is lower than
that of [\cite{bingham94}] taken
in the kinetic regime by a factor $\left(1 - \omega ^2/
\left|{\bf k}\right|^2c^2\right)^2$,
where $\omega $ and ${\bf k}$ 
are the wave frequency and wavenumber respectively, 
and $c$ is the speed of light.  This is an important point, since for plasmas
where the electrons are relativistic or quasi-relativistic, 
$\omega \simeq \left|{\bf k}\right|c$, and this extra factor
gives a big reduction in the calculated growth rate. Below we will present a
plasma theory treatment of the instability that reproduces the quantum
statistical field theory result in the kinetic limit, and gives a result
similar to [\cite{bingham94}] in the reactive limit.

Another important caveat to these works (also emphasised by [\cite{hardy97}]) is
the collisional damping as electrons within the wave collide with ions. Even
the most optimistic calculation gives a neutrino growth rate that is orders
of magnitude lower than the collisional damping rates. Assuming that the wave
mode actually exists, neutrino induced oscillations will be rapidly damped
giving rise to plasma heating. In [\cite{bingham94,hardy97}] the plasma was
assumed to be hydrogenic, at an electron temperature $T_e = 3\times 10^9$ K
and density $n_e = 10^{30}$ cm$^{-3}$, giving of order $\Lambda \sim 100$ 
electrons per
Debye sphere. However heavy elements are often the dominant component of
supernova plasma. For example Cas A is thought to be mainly O 
[\cite{vink96,arnett96}].
In a heavy element plasma $\Lambda $ scales
as $Z^{-3/2}$ where $Z$ is the ion charge. For fully ionized O, $\Lambda\sim
10$, and the so-called plasma approximation may be expected to break down.
In this case no collective motions will exist, and no collisionless energy
deposition can occur.

For this reason this paper will concentrate on degenerate electron plasma 
found at the supernova core. At densities $10^{34} - 10^{38}$ cm$^{-3}$ the
electrons are highly degenerate (Fermi energies 10 - 300 MeV compared with
$kT_e\sim 0.3$ MeV) and collisional
damping is strongly suppressed. The anisotropy in the neutrino distribution
will be taken to to come from the natal kick that a pulsar may acquire during
the supernova event. Various authors ([\cite{horowitz98,lai98a}] and references
therein) have considered the role of neutrino
scattering from partially polarized neutrons in a $10^{15}$ G magnetic field
that may be present as an explanation for this kick, (due to parity violation,
the neutrinos preferentially scatter to directions along the magnetic field).
However a recent paper (which also gives a convenient review of the
observations of pulsar kicks) [\cite{lai98b}] suggests
that this is unlikely to be the actual mechanism. However it is clear that
if a pulsar is born moving with a particular velocity with respect to the
supernova remnant, no matter what the precise mechanism for this is, then the 
neutrino emission must be anisotropic.
This comes about simply because the momentum carried off by the visible 
supernova/supernova 
remnant is generally much smaller than that of the pulsar, and consequently
cannot conserve momentum by itself.
Such a scenario might plausibly result from an off-centre
detonation. For a pulsar moving with velocity 1000 km sec$^{-1}$ an anisotropy
of order 1\% [\cite{chugai84}] must be present in the neutrino emission.

\section{The Neutrino-Plasma Interaction}
The work of [\cite{bingham94}] treated the neutrino plasma interaction by an
effective potential $\sqrt{2}G_Fn_e$, following Bethe [\cite{bethe86}], who was 
considering the neutrino refractive index in the solar core and envelope, and
hence was dealing with a non-relativistic plasma. For neutrino interactions
with a relativistic plasma the effective potential should be generalized. 
The amplitudes $F_W$, $F_Z$ 
for neutrino-electron forward scattering for $W$ and $Z$
exchange are (in units with $\hbar = c = 1$)
[\cite{commins83,itzykson85}] 
\begin{eqnarray}
F_W&=& {-G_F\over\sqrt{2}}\bar{u}_{\nu}
\gamma _{\mu}\left(1-\gamma _5\right)u_{\nu}
\bar{u}_e\gamma ^{\mu}\left(1-\gamma _5\right)u_e\\
F_Z&=& {-G_F\over\sqrt{2}}\bar{u}_{\nu}
\gamma _{\mu}\left(1-\gamma _5\right)u_{\nu}
\bar{u}_e\gamma ^{\mu}\left({1\over 2} -2\sin ^2\theta _W -{1\over 2}
\gamma _5\right)u_e,
\label{e0}
\end{eqnarray}
where $\bar{u}_a$, $u_a$ are final and initial wavefunctions for particles
of species $a$, ($m_a$ their mass, $p_a^{\mu}$, $p_{a,\mu}$ their 4-momenta,)
and $\sin ^2\theta _W \simeq 0.23$ is the Weinberg mixing angle. A Fierz
transformation has been performed in the expression for $F_W$. Only electron
neutrinos interact via $W$ exchange, whereas all three neutrino flavours
interact with electrons through $Z$ exchange. The axial vector terms do not
contribute to the excitation of longitudinal waves, so $F_W>>F_Z$. Henceforward
we consider only $F_W$, putting
\begin{equation}\bar{u}_{\nu}\gamma _{\mu}\left(1-\gamma _5\right)u_{\nu} =
2\bar{u}_{\nu}\gamma _{\mu}u_{\nu} = 2\bar{u}_{\nu}u_{\nu}p_{\nu,\mu}/m_{\nu} 
\end{equation} where
the last term follows from the Gordon identity. For electrons the same 
treament gives $\bar{u}_eu_e\left(p_e^{\mu}/m_e -\mbox{\boldmath $\sigma$}
\right)$, where
$\mbox{\boldmath $\sigma$}$ is a 3-vector composed of the Pauli spin matrices.
Including 
wavefunction normalization factors, $\sqrt{m_a/E_a}$ and
and integrating over a electron distribution
$f_e\left(\bf p_e\right)$ (${\bf p_a}$ is the 3-momentum for particle species
$a$) with no net polarization we get the following 
result for the effective
potential for electron neutrinos
\begin{equation}
V_{eff} = \sqrt{2}G_F\int f_e\left(\bf p_e\right)
\left(1-{{\bf p_{\nu}}\cdot{\bf p_e}\over E_{\nu}E_e}\right) d^3{\bf p_e}. 
\label{e1}
\end{equation} 
Obviously for non-relativistic
plasma the Bethe result is recovered.
Also for isotropic plasma, electrons with oppositely directed
momenta will average to zero in equation (\ref{e1})
recovering $\sqrt{2}G_Fn_e$. However if a plasmon exists in the degenerate
plasma, it must be carried by electrons moving at or very close to the Fermi
velocity, and this scalar product will not necessarily average out to zero.

Following the development and notation 
in [\cite{melrose86}] (pages 19-20) we estimate instability growth
rates.  
The force $F$ acting on an electron becomes
\begin{equation}
F=-q\left({\bf E} + {\bf v_e}\times{\bf B}\right) -\sqrt{2}G_F\nabla
\int f_{\nu}\left({\bf p_{\nu}}\right)\left(1-{{\bf p_{\nu}}\cdot{\bf p_e}\over
E_{\nu}E_e}\right)d^3{\bf p_{\nu}}
\label{e2}
\end{equation} where ${\bf E}$ and ${\bf B}$ 
are electric and magnetic fields, $\omega$ and ${\bf k}$ are the frequency and
wavevector of oscillating quantities, 
$-q$ is the
electron charge and ${\bf v_e}$ the electron velocity. This gives an 
extra current density $J_i$
due to the neutrino interaction
\begin{equation}
J_i=\sqrt{2}qG_F\int\nabla
\int f_{\nu}\left({\bf p_{\nu}}\right)\left(1-{{\bf p_{\nu}}\cdot{\bf p_e}\over
E_{\nu}E_e}\right)d^3{\bf p_{\nu}}{{\bf k}\cdot\partial f_e/\partial{\bf p_e}
\over \omega-{\bf k}\cdot{\bf v_e}}v_i d^3{\bf p_e}.
\label{e3}
\end{equation} Putting $\nabla f_{\nu} =
\nabla \delta f_{\nu} = i{\bf k}\delta 
f_{\nu}$, ${\bf p_a}/E_a ={\bf v_a}$
and \begin{equation}
\delta f_{\nu}\left(\bf p_{\nu}\right) = -\sqrt{2}G_F\int\delta f_e
\left({\bf p_e}\right)\left(1-{\bf v_{\nu}}\cdot{\bf v_e}
\right)d^3{\bf p_e}{{\bf k}\cdot\partial f_{\nu}/
\partial{\bf p_{\nu}}
\over \omega-{\bf k}\cdot{\bf v_{\nu}}},
\label{e4}
\end{equation}
with $J_i=\sigma _{ij}E_j$ and $K_{ij} = 1 + i\sigma_{ij}/\omega $ we arrive
at an expression for the extra term in the dielectric tensor 
$K_{ij}$ due to the neutrino 
interaction;
\begin{eqnarray}
\nonumber\delta K_{ij}=&{2G_F^2q^2\over\omega ^2}\int\int\int
\left\{{\left(\omega -{\bf k}
\cdot{\bf v^{\prime}_e}\right)\delta_{sj}+k_sv^{\prime}_{ej}\over
\omega-{\bf k}\cdot{\bf v^{\prime}_e}}\right\}
{\partial f_e\over\partial {\bf p^{\prime}_e}}\left(1-{\bf v_{\nu}}\cdot
{\bf v^{\prime}_e}\right)d^3{\bf p^{\prime}_e}\\
&\times {{\bf k}\cdot\partial f_{\nu}/
\partial {\bf p_{\nu}}\over \omega -{\bf k}\cdot{\bf v_{\nu}}}\left(1-
{\bf v_{\nu}}\cdot{\bf v_e}\right)d^3{\bf p_{\nu}}{{\bf k}\cdot\partial f_e/
\partial{\bf p_e}\over\omega -{\bf k}\cdot{\bf v_e}}v_id^3{\bf p_e}.
\label{e5}
\end{eqnarray}
Specializing to longitudinal waves, 
\begin{eqnarray}
\nonumber\delta K^L=&{2G_F^2q^2\over\omega ^2}\int\int\int
\left\{{{\bf k}\over\left|{\bf k}\right|}\cdot{\partial f_e\over\partial{\bf
p_e^{\prime}}}{\omega\over\omega -{\bf k}\cdot{\bf v_e^{\prime}}}\right\}
\left(1-{\bf v_{\nu}}\cdot
{\bf v^{\prime}_e}\right)d^3{\bf p^{\prime}_e}\\
&\times {{\bf k}\cdot\partial f_{\nu}/
\partial {\bf p_{\nu}}\over \omega -{\bf k}\cdot{\bf v_{\nu}}}\left(1-
{\bf v_{\nu}}\cdot{\bf v_e}\right)d^3{\bf p_{\nu}}
{{\bf k}\cdot{\bf v_e}\over\left|{\bf k}\right|}
{{\bf k}\cdot\partial f_e/
\partial{\bf p_e}\over\omega -{\bf k}\cdot{\bf v_e}}d^3{\bf p_e}.
\end{eqnarray}
This expression is evaluated by noting that integrals of the form
$\int{\bf v_e}{{\bf k}\cdot\partial f_e
/\partial {\bf p_e}\over\omega -{\bf k}\cdot{\bf v_e}}d^3{\bf p_e}$ are parallel
to ${\bf k}$ for a spherically symmetrial electron distribution function 
$f_e\left({\bf p_e}\right)$, and those
$\int{\bf v_e}{{\bf k}\cdot\partial f_e
/\partial {\bf p_e}\over\omega }d^3{\bf p_e}=0$.
Hence we may put \begin{equation}{\bf v_{\nu}}\cdot{\bf v_e} 
\rightarrow
{\bf v_{\nu}}\cdot{\bf k}~{\bf v_e}\cdot{\bf k}/\left|\bf k\right|^2
\end{equation} and 
perform the integration using the standard results in Table 1,
putting $\partial f_e/\partial\left|{\bf p_e}\right| = -2/\left(2\pi \right)^3 
\times\delta\left(\left|{\bf p_e}\right| - p_F\right)$
and $\omega _p^2 = n_eq^2/\gamma _Fm_e$. We have neglected
the variation of the relativistic $\gamma$ inside the integral and have
approximated
it by its value at the Fermi surface.
The full result is
\begin{equation}
K^L=1+{3\over 2}{\omega _p^2\over \left|{\bf k}\right| ^2v_F^2}
\left(2-{\omega\over \left|{\bf k}\right| v_F}\log\left(
\omega +\left|{\bf k}\right| v_F\over \omega -\left|{\bf k}\right|
v_F\right)\right)\left(1+\Delta\right)
\label{e24}
\end{equation}with
\begin{eqnarray}
\nonumber\Delta=&2G_F^2\int\left(1-{{\bf k}\cdot
{\bf v_{\nu}}\over\left|{\bf k}\right| }{\omega\over \left|{\bf k}\right| }
\right) ^2{3\over 2}{n_e\over p_Fv_F}
\left(2-{\omega\over\left|{\bf k}\right| v_F}\log\left(\omega +\left|{\bf k}
\right| v_F\over\omega -\left|{\bf k}\right| v_F\right)\right)\\
&\times
{{\bf k}\cdot\partial f_{\nu}/\partial {\bf p_{\nu}}\over \omega -{\bf k}\cdot
{\bf v_{\nu}}} d^3{\bf p_{\nu}}.
\label{e7}
\end{eqnarray}
and the subscript $F$ denoting values of variables at the Fermi surface.
If we let $\Delta\rightarrow 0$ we recover the usual plasmon dispersion relation
for degenerate electrons (cf [\cite{braaten93}]). This dispersion relation is
plotted in Figure 1 for an electron density of $10^{34}$ cm$^{-3}$. The vertical
dotted line indicates the wavevector where $\omega = \omega _{crit} =
\left|\bf k\right|$. To
the right of this line $\omega < \left|\bf k\right|$ and neutrinos may
excite electrostatic waves by a kinetic instability. To the left of this line
they are kinematically forbidden from doing so, and only the reactive 
instability may operate. This line moves to higher wavenumber
(in units of the  plasma frequency) with increasing electron density.

\begin{table}
\caption{Standard Integrals}
\begin{tabular}{ll}
$\int _1^1 {kvx\over \omega -kvx}dx=$&  $-2+{\omega\over kv}\log\left({\omega
+kv\over\omega -kv}\right)$ \\
$\int _1^1 {kvx^2\over \omega -kvx}dx=$&  
${\omega\over kv}\left(-2+{\omega\over kv}\log\left({\omega
+kv\over\omega -kv}\right)\right)$ \\
$\int _1^1 {kvx^3\over \omega -kvx}dx=$& $ 
-{2\over 3} +\left({\omega\over kv}\right)^2
\left(-2+{\omega\over kv}\log\left({\omega
+kv\over\omega -kv}\right)\right)$ \\
$\int _1^1 {kvx^4\over \omega -kvx}dx=$&$  
{\omega\over kv}\left(-{2\over 3} +\left({\omega\over kv}\right)^2
\left(-2+{\omega\over kv}\log\left({\omega
+kv\over\omega -kv}\right)\right)\right)$ \\
\end{tabular}
\end{table}

\subsection{The Kinetic Growth Rate}
To derive the instability growth rate in the kinetic regime we let
$\omega\rightarrow\omega + i\delta$ with $\omega >> \delta$ and put
${1\over \omega-{\bf k}\cdot{\bf v_{\nu}}}\rightarrow
P\left(1\over \omega-{\bf k}\cdot{\bf v_{\nu}}\right) -i\pi\delta\left(
\omega -{\bf k}\cdot{\bf v_{\nu}}\right)$, which gives, after taking
imaginary parts,
\begin{eqnarray}
\delta =& {2\pi G^2_F\omega\left|{\bf k}\right| ^2\over q^2}\left(\omega^2 
-\left|{\bf k}\right| ^2v_F^2\over
3\omega _p^2 -\left(\omega^2 -\left|{\bf k}\right| ^2v_F^2\right)\right)
\left(1-{\omega ^2
\over\left|{\bf k}\right| ^2}\right)^2\\
&\times\int {\bf k}\cdot{\partial f_{\nu}\over\partial {\bf 
p_{\nu}}}\delta\left(\omega -{\bf k}\cdot{\bf v_{\nu}}\right) d^3{\bf p_{\nu}}.
\label{e8}
\end{eqnarray}
This result is identical to that derived in [\cite{hardy97}] by the methods of
quantum statistical field theory, with the substitution of the appropriate
residue factor for degenerate electrons [\cite{braaten93}]
\begin{equation}
Z\left(\left|{\bf k}\right|\right)=\left[1+{1\over 2\omega}{\partial\alpha 
\over\partial\omega}\right]^{-1}
_{\omega =\omega\left(\left|{\bf k}\right|\right)}
={2\left(\omega ^2 -\left|{\bf k}\right| ^2v_F^2\right)\over 3\omega _p^2 -
\left(\omega ^2 -\left|{\bf k}\right| ^2v_F^2\right)}
\end{equation}
where $\omega ^2 = -\alpha\left(\omega\right)$ is the plasmon dispersion
relation, and we have substituted from the dispersion relation to eliminate
the logartithm. Similar results for spontaneous Cherenkov emission by 
neutrinos have been obtained by [\cite{d'olivo96,sawyer92}].
It is apparent the the factor $\left(
1-\omega ^2/\left|{\bf k}\right| ^2\right)^2$ 
by which this expression differs from the
kinetic limit of the growth rate derived by [\cite{bingham94}] is not related
to electro-weak parity violation effects as implied by [\cite{hardy97}], but
to a proper treatment of the relativistic plasma.
In fact it is straightfoward to redo the calculation of
the neutrino-plasmon vertex in [\cite{hardy97}] admitting right-handed as well 
as left-handed neutrinos (i.e. replacing $\bar{u}_{\nu}\gamma _{\mu}
\left(1-\gamma _5
\right)u_{\nu}$ with $2\bar{u}_{\nu}\gamma _{\mu}u_{\nu}$) 
The result is the same as for the pure 
left-handed neutrinos
except for an extra factor of 2.

\subsection{The Reactive Growth Rate}
Turning to the reactive instability, considered in [\cite{bingham94}], where
the pole in the integrand lies in a region where $\partial f_{\nu}/\partial
{\bf p_{\nu}} =0$. This will be the case in the region of phase space where
the plasmon dispersion relation satisfies $\omega >> 
\left|\bf k\right|$.
Using the dispersion relation to eliminate the $\log$
inside the integral and putting
$2-{\omega\over\left|{\bf k}\right| v_F}\log\left(\omega +\left|{\bf k}\right|
v_F\over \omega -\left|{\bf k}\right| v_F\right)
=-{2\over 3}{\left|{\bf k}\right| ^2v_F^2\over\omega ^2}
-{2\over 5}{\left|{\bf k}\right| ^4v_F^4\over\omega ^4}-\cdots$ for $\omega
>> \left|\bf k\right|$ elsewhere, we solve equations (\ref{e24}) and (\ref{e7})
with $K^L=0$ for to get
\begin{eqnarray}\nonumber
&\omega ^2 -\omega _p^2\left(1+{3\over 5}{\left|\bf k\right|^2v_F^2\over
\omega ^2}\cdots + \right)\\
&\times\left\{1-{2G_F^2\left|\bf k\right|^2\over q^2}
\int\left(1-{{\bf k}\cdot{\bf v_{\nu}}\over\left|\bf k\right|}{\omega\over
\left|\bf k\right|}\right) ^2
{{\bf k}\cdot\partial f_{\nu}/
\partial
{\bf p_{\nu}}\over \omega -{\bf k}\cdot{\bf v_{\nu}}}d^3{\bf p_{\nu}}\right\} =0
\label{e9}
\end{eqnarray}
This is similar to the expression derived in [\cite{bingham94}], but different
by the factor
$\left(1-{{\bf k}\cdot{\bf v_{\nu}}\over\left|\bf k\right|}{\omega\over
\left|\bf k\right|}\right) ^2$
in the integral. The integral in equation (\ref{e9}) can be further simplified
by taking the limit $\omega >> \left|\bf k\right|$.
The reactive growth rate is
given by the complex roots (if any) to equation (\ref{e9}). Assuming that
$\int {\bf k}\cdot\partial f_{\nu}/\partial {\bf p_{\nu}} d^3{\bf p_{\nu}}=0$
(this is true for the distribution functions considered below), then to 
lowest order
in $\left|\bf k\right|/\omega$  a quadratic
equation in $\omega$ with two real roots results, giving no reactive growth.
Going to higher order a
cubic equation in $\omega ^2$ results, again with no complex roots. A negative
real root for $\omega ^2$ would give an imaginary $\omega$, implying reactive
growth at zero frequency. However such solutions should not be 
over-interpreted. At low frequencies electrons will couple to protons, which in
turn will couple to neutrons by the strong interaction, and these couplings
would need to be included in the dispersion relation to get a physical result.

\section{The Neutrino Distribution Function and Numerical Estimates}
To get numerical estimates of the magnitudes of the various growth rates we
will consider a neutrino distribution function of the form
\begin{equation}
f_{\nu}\left({\bf p_{\nu}}\right)={1\over\left(2\pi\right)^3}{1\over\exp\left(
\left|{\bf p_{\nu}}\right|/k_BT_{\nu}\right)+1}\left(1+\tau P_n\cos\theta _{\nu}\right).
\label{e10}
\end{equation}
which comes from a thermal Fermi-Dirac distribution (the first term, with the
neutrino chemical potential assumed zero) modified
by scattering with an approximate differential cross section [\cite{horowitz98}]
\begin{equation}
d\sigma /d\Omega = \sigma_0\left(1+P_n\cos\theta\right)
\end{equation}
with $\sigma_0= G_F^2E_{\nu}^2/4\pi ^2$. The neutrino opacity, $\tau =\tau _0
E_{\nu}^2\sim \left(E_{\nu}/{\rm MeV}\right)^2$ cm$^2$ for typical neutron
star parameters. The neutron polarization $P_n$ is given approximately by
[\cite{horowitz98}]
\begin{equation} P_n=2\times 10^{-5}\left(B/10^{13}{\rm G}\right)\left(3 {\rm
MeV}/T_n\right)\end{equation} where $T_n$ is the temperature of the neutron
star matter. Consequently $\tau P_n <<1$ for reasonable parameter values and
equation(\ref{e10}) is justified.
As remarked above this neutrino distribution function based on elastic 
scattering from partially polarized neutrons is unlikely to be the cause of
pulsar kicks [\cite{lai98b}]. The reason for this is that the anisotropy in
the neutrino distribution is wiped out by neutrino
absorption by neutrons, unless a difference in the magnetic field of at least
$10^{16}$ G exists between the two opposite poles of the newly formed neutron
star. However the neutrino distribution function that results when a pulsar
receives a kick is likely to be of this form, so we proceed assuming
equation (\ref{e10}). For the plasma instabilities, the important feature is
that neutrino scattering cross section {\it rise with energy} as $E_{\nu}^2$.
Thus the scattered (i.e. anisotropic)  part of the neutrino distribution
function will have a positive value of $\partial f_{\nu}/\partial p_{\nu}$
giving rise to postive kinetic growth rates in this region of phase space.

\subsection{The Kinetic Case}
The integral over the neutrino distribution function may be carried out
putting \begin{equation}
{\bf k}\cdot\partial f_{\nu}/\partial {\bf p_{\nu}} =\left|{\bf k}\right|
\cos\beta\partial f_{\nu}/\partial\left|{\bf p_{\nu}}\right| -\left|{\bf k}
\right| \sin\beta /\left|{\bf p_{\nu}}\right| \partial f_{\nu}/
\partial\theta _{\nu} 
\label{e22}
\end{equation}
with $\cos\beta = {\bf k}\cdot{\bf p_{\nu}}/\left|{\bf k}\right|\left|{\bf
p_{\nu}}\right|$. For emission along $\theta _{\nu}=0$, $\beta =-\theta _{\nu}$
and using the standard result $\int _0^{\infty}x^{n-1}/\left(\exp\left(x\right)
+1\right)dx =\Gamma\left(n\right)\zeta\left(n\right)\left(1-1/2^{n-1}\right)$
the integral evaluates to 
\begin{equation}
\int {\bf k}\cdot{\partial f_{\nu}\over\partial {\bf 
p_{\nu}}}\delta\left(\omega -{\bf k}\cdot{\bf v_{\nu}}\right) d^3{\bf p_{\nu}}
= -{\left(k_BT_{\nu}\right) ^2\over 24} 
+{7\pi ^2\over 240}
\left(k_BT_{\nu}\right) 
^4\tau _0P_n
\label{e23}
\end{equation} for $\omega\sim\left|{\bf k}\right|$. The first
term gives a negative contribution due to neutrino Landau damping by the
thermal part of the distribution function. The second term gives rise to the
instability. Thus $0.7\pi ^2\left(k_BT_{\nu}\right)^2\tau _0P_n>1$ is required.
This evaluates to $k_BT_{\nu}\sim 85$ MeV, comparable to the gravitational
binding energy per nucleon released in the collapse, for 
$P_N\sim 2\times 10^{-5}$. At higher magnetic field strengths such as those
thought to be present in magnetars [\cite{kouveliotou98,vasisht97}] 
($B\sim 10^{15} G$) the required
neutrino temperature drops to 8.5 MeV. This temperature (though probably not the
magnetic field) is more consistent with neutrinos actually
observed from SN 1987A [\cite{bahcall89}].

However as pointed out by [\cite{hardy97}] the kinetic growth rate will still
be very small. The reason is that neutrinos may only excite waves with phase
velocities in the range $v_F\rightarrow 1$. Reinstating the necessary factor
of $\hbar ^{-3}c^{-5}$ in equation(\ref{e8}) and evaluating gives an approximate
expression for the maximum growth rate\begin{equation}
\delta\sim 1.8\times 10^{-53}{n_e\omega _p \over\gamma _F}\left(1-v_F\right)^3.
\label{e25}
\end{equation} for a neutrino temperature of 10 MeV, and the value of the
integral in equation (\ref{e23}) taken to be $+\left(k_BT_{\nu}\right)^2/24$.
This equation is evaluated for a variety of densities in Table
2. These values are far too small to be important during
the few second neutrino burst during a supernova explosion. The lower densities
in Table 2 may occur outside the neutrinosphere (the central
region where neutrinos are trapped and are essentially isotropic, see 
[\cite{arnett96}]). In this case
a beam instability of the type considered in [\cite{bingham94,hardy97}] would
be more realistic.
\begin{table}
\caption{Kinetic Instability Parameters}
\begin{tabular}{llllll}
$n_e$ (cm$^{-3}$)& $1-v_F/c$& $\gamma $& $\omega _p$ (rad/s)& $\omega _{crit}$
($\omega _p$)& $\delta$ (rad/s)\\
$10^{32}$& 0.016& 5.6& $2.4\times 10^{20}$& 2.11& $3\times 10^{-7}$ \\
$10^{34}$& $7.5\times 10^{-4}$& 26& $1.1\times 10^{21}$& 2.92& 
  $3\times 10^{-9}$ \\
$10^{36}$& $3.2\times 10^{-5}$& 120& $5.2\times 10^{21}$& 3.64& 
  $3\times 10^{-11}$ \\
$10^{38}$& $1.6\times 10^{-6}$& 550& $2.4\times 10^{22}$& 4.21& 
  $3\times 10^{-13}$ \\
\end{tabular}
\end{table}

\subsection{The Reactive Case}
With the same notation as above, the integral over the second term in
equation (\ref{e22}) gives zero, while the first term evaluates to
$\left(k_BT\right)^2/36$. As expected this term gives no wave growth, since
it represents the Landau damping of waves by the thermal neutrino distribution.
We remark here that our analysis of the reactive instability is necessarily
rather superficial. It would not be difficult to assume a neutrino distribution
function that would give wave growth. However the question of whether such
a distribution function is realistic for a supernova would still not be 
answered. For this reason the subsequent discussion will be concerned only
with the kinetic instability.

\section{Discussion and Conclusions}
The fundamental reason that keeps the kinetic instability from developing
significant wave growth is that the phase velocity of the waves supported
by the plasma is very close to unity (the speed of light). Hence the factor
of $\left(1-\omega ^2/\left|\bf k\right| ^2\right)^2$ in the growth rate becomes
very small. The physical reason for this is that when the dispersive properties
of the plasma approach those of the vacuum, longitudinal waves will cease to
exist. With this in mind we will briefly consider the effect of the magnetic
field on the plasmon dispersion relation.
Again working from ref [\cite{melrose86}] (page 166, equation 10.21) we can 
derive an expression for $K^L$ with an ambient magnetic field. For simplicity
we consider the case ${\bf k}\cdot {\bf B} =0$. For ${\bf k}\Vert {\bf B}$,
the problem 
is identical to the non-magnetic case discussed above.
Putting 
$\partial f_e/\partial p_{e\perp} = 
-2/\left(2\pi \right)^3 \times\delta\left(p_{e\perp} -\sqrt{p_F^2
 - p_{e\Vert}^2}\right)$,
$J_{s-1}\left(x\right) + J_{s+1}\left(x\right) = {2s\over x}J_s\left(x\right)$,
and rewriting the sum over Bessel functions as 
$\sum _{\infty}^{\infty}\rightarrow
\sum _0^{\infty}$ the expression is
\begin{equation}
K^L=1-\int _0^{\pi}
\sum _{s=0}^{\infty}{3\omega _ps^2\over \omega ^2 -s^2\Omega ^2}
J_s^2\left({\left|
{\bf k}\right| v_F\sin\theta\over\Omega}\right){\Omega^2\over\left|{\bf k}
\right| ^2v_F^2}\sin\theta d\theta .
\end{equation}
The electron cyclotron frequency is 
$\Omega = q\left|\bf B\right|/\gamma _F m_e$.
For $\Omega << \left|\bf k\right|v_F$ the sum over $s$ can be replaced by an 
integral, whereupon its evaluation recovers the non-magnetic longitudinal
dispersion relation. It is more interesting to consider the behaviour of
individual harmonics over the range 
$\left|\bf k\right|v_F<<\Omega\rightarrow 
\left|\bf k\right|v_F>>\Omega$. Then a behaviour something like the Bernstein 
modes discussed in [\cite{melrose86}] appears. For example for $s=1$, as
$\left|\bf k\right|\rightarrow 0$, $\omega\rightarrow\Omega _{UH}$ where
$\Omega _{UH}^2=\omega _p^2 +\Omega ^2$ is the upper hybrid frequency. However
as $\left|\bf k\right|\rightarrow\infty $, $\omega\rightarrow\Omega\ne
\left|\bf k\right|$, so a kinetic instability could have significantly higher
growth rate than in the estimates above. Another possibility if $\Omega > 
\omega _p$ is a kinetic instability exciting electromagnetic waves, since
for low frequencies $K_T = \left|\bf k\right|^2/\omega ^2 > 1$. The discovery
of pulsars with magnetic fields $\sim 10^{15}$ G 
[\cite{kouveliotou98,vasisht97}]
opens up this possibility 
for degenerate electrons at densities $10^{33} - 10^{34}$ cm$^{-3}$. These
novel scenarios will be investigated further in future papers.

In conclusion then
we have given a plasma theory for the so-called neutrino instability, that
reproduces the results of quantum statistical field theory in the kinetic
limit, and which gives more accurate expressions to use in the development
of the reactive instability. The growth rates for longitudinal waves are
insignificant in, e.g. the neutrino nurst during a supernova explosion, but
we have been able to identify plausible conditions in which such wave growth
might occur.

\begin{figure}
\plotone{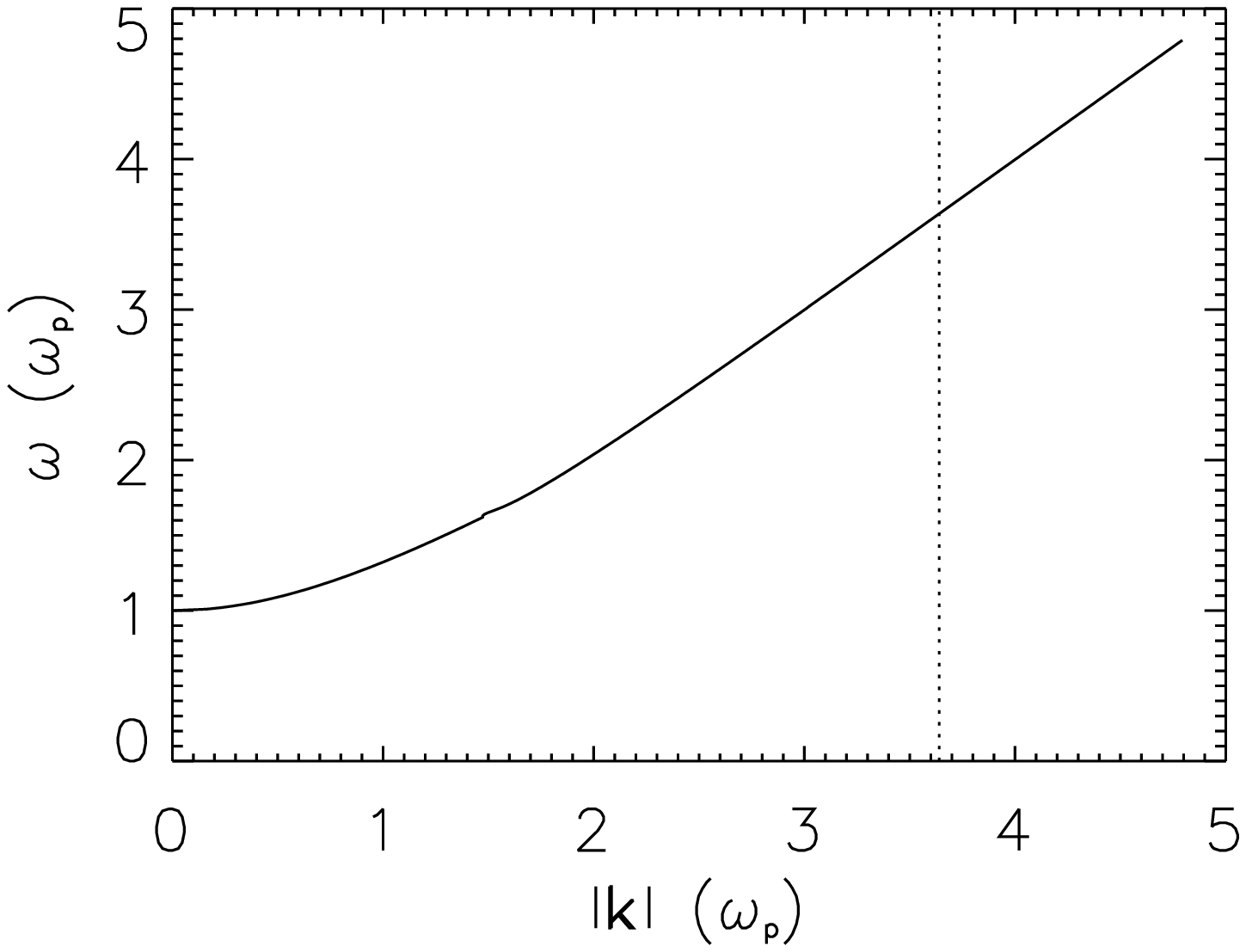}
\caption{The dispersion relation for electrostatic waves in a degenerate
electron plasma of density $10^{34}$ cm$^{-3}$. The vertical dotted line
indicates the wavevector where $\omega =\left|{\bf k}\right|$, which is the
onset of the kinetic instability regime.}
\end{figure}
\end{document}